\documentclass[conference]{IEEEtran}
\IEEEoverridecommandlockouts
\usepackage{cite}
\usepackage{amsmath,amssymb,amsfonts}
\usepackage{algorithmic}
\usepackage{graphicx}
\usepackage{textcomp}
\usepackage{xcolor}
\usepackage{subfig}
\usepackage{booktabs}
\usepackage{import}
\usepackage{color}
\usepackage[capitalise]{cleveref}
\usepackage{multirow}
\usepackage{array}
\usepackage[font=footnotesize]{caption}
\usepackage[inline]{enumitem}
\usepackage{makecell}
\usepackage{siunitx}
\usepackage{caption}
\usepackage[acronym,toc]{glossaries}
\glsdisablehyper
\makeglossaries
\newacronym{drl}{DRL}{deep reinforcement learning}
\newacronym{madrl}{MA-DRL}{multi-agent deep reinforcement learning}
\newacronym{sarsa}{SARSA}{state–action–reward–state–action}
\newacronym{leo}{LEO}{low Earth orbit}
\newacronym{qos}{QoS}{quality of service}
\newacronym{geo}{GEO}{geostationary orbit}
\newacronym{meo}{MEO}{medium Earth orbit}

\newcommand{\showmlcomments}{}
\ifdefined\showmlcomments
    \newcommand{\oldtext}[1]{\textcolor{teal}{#1}}
    \newcommand{\ml}[1]{\textcolor{olive}{#1 }} 
    \newcommand{\mlr}[1]{\textcolor{green!60!black}{[ML: #1]}}
\else
    \newcommand{\oldtext}[1]{}
    \newcommand{\ml}[1]{}
    \newcommand{\mlr}[1]{}

\fi

\usepackage[top=0.7in, bottom=1.05in, left=0.7in, right=0.7in]{geometry}
\begin{document}
\bstctlcite{BSTcontrol}
%
\title{Queue-Aware and Resilient Routing in LEO Satellite Networks Using Multi-Agent Reinforcement Learning}
%
%
%

\author{\IEEEauthorblockN{Mudassar~Liaq, Mahyar Tajeri and Peng Hu}
\IEEEauthorblockA{Department of Electrical and Computer Engineering, University of Manitoba, MB, Canada.\\
  \{mudassar.liaq, mahyar.tajeri,  peng.hu\}@umanitoba.ca \\ }
\thanks{We acknowledge the support provided by the Government of Canada and Natural Sciences and Engineering Research Council of Canada (NSERC), [funding reference number RGPIN-2022-03364].}}

%
%

\markboth{Journal of \LaTeX\ Class Files,~Vol.~14, No.~8, August~2015}%
{Shell \MakeLowercase{\textit{et al.}}: Bare Demo of IEEEtran.cls for IEEE Journals}
%



\maketitle

\begin{abstract}
With the rapid growth in data demand and stringent latency requirements of modern applications has driven significant interest in Low Earth Orbit (LEO) satellite constellations as an emerging solution for global Internet coverage. However, routing in LEO networks remains a fundamental challenge due to highly dynamic topologies, time-varying traffic conditions, and its susceptibility to link failures. Conventional routing algorithms typically assume static link metrics and fail to account for queue backlogs or real-time system variations, making them less effective in such environments.
We propose a queue-aware multi-agent deep reinforcement learning (MA-DRL) framework for routing in LEO satellite networks. Each satellite is modeled as an independent agent responsible for making local routing decisions, enabling a distributed and scalable solution. The proposed framework formulates a latency-aware optimization problem that incorporates background traffic, queue dynamics at each satellite, and a resilience score to improve robustness.
We evaluate the proposed approach against the state-action-reward-state-action (SARSA) and Dijkstra algorithms. While Dijkstra achieves the lowest end-to-end latency under ideal conditions, its computational and signaling overhead becomes a significant bottleneck as the network scales. In contrast, our proposed approach incurs significantly lower overhead (approximately 50\% of Dijkstra at a 5 s recalculation interval), scales efficiently with network size, and effectively manages queue backlogs and resilience under increasing traffic load, demonstrating enhanced robustness and scalability in LEO satellite networks while maintaining competitive latency and resilience scores.
\end{abstract}



\IEEEpeerreviewmaketitle

\glsresetall
\section{Introduction}
%
%
As the demand for low-latency and data-intensive applications continues to grow, \gls{leo} satellite networks are increasingly being utilized to provide global internet coverage \cite{intro_1_handley2018delay}. While these \gls{leo} satellites enable connectivity for data and latency sensitive applications at a global scale, they also introduce several fundamental challenges. In particular, the rapidly changing topology, intermittent connectivity, and large-scale deployment of satellites make efficient and optimized routing a complex problem \cite{intro_2_del2019technical}. Traditional routing approaches often fail to capture these dynamics and are therefore not well suited for \gls{leo} constellations.

Various works have explored routing and resource optimization to enhance network performance under such dynamic conditions, with a focus on latency, power consumption, and time-varying link characteristics\cite{2__liang2023free,4__han2025multi, 5__li2025performance}. Some studies have addressed \gls{qos} requirements by integrating higher-orbit satellite systems, such as \gls{geo} and \gls{meo}, to improve service continuity and network resilience.  Additionally, while some works \cite{multipath_1_rai2025optimized} aim to improve network resilience through secondary routing paths or multi-orbit integration, significant research gaps still exist in jointly addressing latency, congestion, and robustness. More recently, machine learning-based solutions, particularly \gls{drl} approaches, have been investigated to enable adaptive and distributed routing decisions \cite{1_2__soret2024q, 1__lozano2025continual}.

Despite these advancements, several challenges are still remaining. Many existing approaches rely on simplified assumptions and fail to capture the practical complexities of \gls{leo} satellite systems\cite{2__liang2023free, 5__li2025performance}. Although \gls{drl}-based methods have been widely explored, most focus primarily on latency minimization without explicitly considering network queue conditions, which play a critical role in such environments, especially under varying traffic loads. To address this challenge, this work makes the following contributions: (i) We propose a queue-aware multi-agent deep reinforcement learning (MA-DRL) framework for routing in LEO satellite constellations, where each satellite makes decentralized routing decisions based on local and neighboring queue information. (ii) We incorporate network resilience into the routing framework by focusing on reducing the probability of node failures, rather than solely relying on post-failure recovery mechanisms. (iii) We evaluate the proposed approach against well-established baseline methods, including Dijkstra and \gls{sarsa}, providing a comprehensive performance comparison. (iv) We demonstrate that the proposed framework achieves improved scalability and adaptability in dynamic LEO environments, effectively balancing latency, congestion, and network stability.


The remainder of the paper is structured as follows. Section II discusses the related work. Section III presents the proposed system model. Section IV discusses the proposed solution approach, followed by the performance evaluation in Section V. Section VI outlines the conclusive remarks.

\section{Related Work}

Recent works in \gls{leo} satellite networks have explored routing, resource optimization, and system design under dynamic and time-varying topologies.
Several studies focused on routing and performance analysis. In~\cite{2__liang2023free}, the trade-off between inter-satellite distance, hop count, transmission power, and latency was analyzed using shortest-path routing in Starlink and Kuiper constellations. Similarly,~\cite{5__li2025performance} investigated the impact of dynamic link topology and channel conditions on downlink performance using detailed physical-layer modeling. 
\Gls{qos} provisioning has also been studied in multi-layer satellite systems. In~\cite{4__han2025multi}, a joint \gls{geo}--\gls{leo} architecture was proposed to improve service continuity, with optimization formulated to maximize user rate and system throughput. Similarly,~\cite{7__sun2025visibility} addressed multi-orbit satellite selection as a joint optimization problem under power and connectivity constraints.
 
Hybrid and integrated architectures have been explored to enhance system capabilities. In~\cite{8__lee2022integrating}, a UAV--\gls{leo} system was proposed, where \gls{drl} was used to optimize trajectory and communication decisions. Similarly,~\cite{9__al2025collaborative} investigated multi-band and multi-orbit integration for efficient resource utilization across heterogeneous links.
Resilience in \gls{leo} networks has also been investigated. In~\cite{3__lai2023achieving}, routing strategies were designed to maintain performance under failures using constraint optimization. Structural and temporal resilience were further analyzed in~\cite{26__zhou2025resilience,27__guo2025resilience} using graph-based models to evaluate network robustness and node importance.
Machine learning-based approaches have also gained attention for routing in LEO constellations. In~\cite{1_2__soret2024q}, a distributed Q-learning framework was proposed for delay minimization using local information, highlighting the impact of queuing delay under congestion. Building on this,~\cite{1__lozano2025continual} introduced a \gls{madrl} framework with federated learning for decentralized routing decisions.

Despite these advancements, several key limitations remain in existing works. Many studies assume simplified conditions and do not explicitly account for dynamic factors such as node failures and time-varying traffic patterns, while also overlooking internal network load conditions, particularly queue dynamics, which play a critical role in congestion and delay. Although integrated architectures, such as joint \gls{leo}--\gls{geo} systems, have been explored, they primarily emphasize structural enhancements rather than adaptive, traffic-aware routing. In addition, the notion of resilience remains inconsistent, with most approaches focusing on post-failure recovery rather than proactive mitigation. Furthermore, while \gls{madrl} has been applied for decentralized routing, existing works provide limited benchmarking and lack comprehensive comparisons with classical and optimization-based baselines.

\section{System Model}
We consider a \gls{leo} satellite constellation composed of $O$ orbital planes, each containing $N_o = \frac{N}{O}$ evenly spaced satellites. The set of orbital planes is denoted by $o \in \{1, 2, \dots, O\}$. Each plane is located at an altitude $h_o$ above the Earth's surface, with an inclination angle $\delta$ and latitude $\epsilon_o$ (in radians). The satellite orbits are assumed to be circular, with a negligible eccentricity (e.g., $e \approx 0.00001$).
Each satellite is equipped with five communication antennas: four for inter-satellite links (ISLs) and one for ground-to-satellite communication. Among the four ISL antennas, two are dedicated to intra-plane communication (i.e., communication with neighboring satellites within the same orbital plane), while the remaining two are used for inter-plane communication (i.e., communication with satellites in adjacent orbital planes)~\cite{1_2_3__leyva2021inter}.
Each satellite $i \in N$ can communicate with all four antennas with other satellites i.e., $|\varepsilon_{i, sat}| \leq 4$. 
On ground, we have gateways $\mathbb{G}$ across Earth, with each represented as $(g_{lat}, g_{lon})$ for $g \in \mathbb{G}$. 
The mobile devices on the ground communicate with the satellite network using these gateways. 
These gateways collect data from connected devices, aggregate it, and then transmit it to their linked device. 
We designate two of these gateways as source and destination ground station, i.e., $s, d \in \mathbb{G}$. 
Each gateway $g \in \mathbb{G}$ keeps one ground-to-satellite link to closest satellite $i \in \mathbb{S}$ within its visible range.
This link is denoted by $\varepsilon_{(i,g)}$, and for all ground terminals, it can be denoted as $\varepsilon_G = \bigcup_{g \in G} \varepsilon_{(i,g)}$. 

We can classify each link $(i,j)$ to one of three categories. A link connecting a gateway node to a satellite node (i.e., $i \in \mathbb{G}$ and $j \in N$) represents an uplink, while a link from a satellite node to a gateway node (i.e., $i \in N$ and $j \in \mathbb{G}$) corresponds to a downlink. When both nodes $i,j \in N$, the link represents an inter-satellite link (ISL).
The overall node set is defined as $\mathbb{N} = N \cup \mathbb{G}$, and the set of communication links can be expressed as $\varepsilon = \varepsilon_N \cup \varepsilon_{\mathbb{G}}$, where $\varepsilon_N$ denotes inter-satellite links and $\varepsilon_{\mathbb{G}}$ denotes ground-to-satellite links.

\subsection{Inter-satellite communication}
The data rate on an inter-satellite link (ISL) can be defined as:

\begin{equation}
    R{(i,j)} = log_2\big( 1 + \frac{P_{(i,j)}d_{(i,j)}^{-\alpha}|h_{i,j}|^2G_iG_j}{q\sigma^2}\big)
    \label{eq_datarate}
\end{equation}
Here, $P_{(i,j)}$ denotes the transmission power of satellite $i$ when communicating with satellite $j$, while $G_i$ and $G_j$ represent the antenna gains of satellites $i$ and $j$, respectively. The distance between the two satellites is denoted by $d_{(i,j)}$, and $\alpha$ represents the path-loss exponent. The term $h_{(i,j)}$ corresponds to the normalized channel gain, and $\sigma^2$ denotes the noise power at the receiver. Furthermore, $q = \frac{4\pi}{\lambda}$ represents the free-space path-loss factor at unit distance, where $\lambda$ is the wavelength.
Based on these parameters, the signal-to-noise ratio (SNR) of the link between satellites $i$ and $j$, denoted by $\mu_{(i,j)}$, is defined as
\begin{equation}
    \mu_{(i,j)} = 2^{R(i,j)/W} - 1
\end{equation}
where $W$ is the channel bandwidth.

\subsection{Space-to-ground communication}
The received SNR over the link between \gls{leo} satellite $i$ and ground base station $j$ can be written  as 

\begin{equation}
    \mu_j^d = \frac{P_{(i,j)}G_{j}G_{i}|g_j^d|^2}{\sigma^2}
\end{equation}

and uplink as: 

\begin{equation}
    \mu_j^u = \frac{P_{(j,i)} G_{j}G_{i}|g_j^u|^2}{\sigma^2} 
\end{equation}
where $P_{(i,j)}$ is the transmission power of the satellite $i$ for the link to the ground station $j$, $G_{i}$ and $G_{j}$ are maximum antenna gains of satellite $i$ and ground station $j$ respectively. 
$g_n^u$ and $g_n^d$ are Nakagami-m distributed uplink and downlink channel coefficients, respectively. 

The per-hop latency incurred when transmitting a packet from the $i^{\text{th}}$ node (satellite) to the $j^{\text{th}}$ node can be decomposed into three main components: queueing delay, transmission delay, and propagation delay. The queueing delay at node $i$, denoted by $t_q(i)$, represents the time a packet spends waiting in the FIFO transmission buffer before it is served. Let $q_i$ denote the queue length at node $i$, and $B$ represent the packet size. Since packets ahead in the queue may be destined for different links with varying transmission rates $R(i,\cdot)$, the queueing delay can be approximated as $t_q(i) = q_i \cdot B / R(i,\cdot).$
The transmission delay corresponds to the time required to place the packet onto the communication link between nodes $i$ and $j$, and is given by $\frac{B}{R(i,j)}$, where $R(i,j)$ denotes the transmission rate of the link. The propagation delay accounts for the time taken by the signal to travel the physical distance between nodes $i$ and $j$. This is given by $\frac{\|ij\|}{c}$, where $\|ij\|$ represents the distance between the two nodes and $c$ is the speed of light.

Accordingly, the total per-hop latency from node $i$ to node $j$ can be expressed as \cite{1__lozano2025continual}:
\begin{equation}
D(i,j) = \frac{\|ij\|}{c} + \frac{B}{R(i,j)} + t_q(i).
\end{equation}

\subsection{Outage probability}
We can define the outage probability for uplink and downlink to be 

\begin{equation}
    P_{out}^{u,d} = \text{Pr}\{\mu_j^u \leq \mu^{th} || \mu_j^d \leq \mu^{th}\}
\end{equation}
where $\mu^{th}$ is the minimum SNR threshold we can tolerate\cite{29__abdulkarim2024multi}. 
The outage probability for all the links in the satellite network can be defined as: 
\begin{equation}
    P^{N}_{out} = \prod_{\forall i,j \in N~\&~i \neq j} (1 - P^{i,j}_{out})~\times~S_{(i,j)}  
\end{equation}
where $S_{(i,j)} \in \{0,1\}$ is the path selection variable, and denotes weather the link is selected as part of optimal path.
$P_{out}^{(i,j)}$ is the outage probability of each hop in the satellite network and can be defined as: 
\begin{equation}
    P_{out}^{i, j} = Pr{\frac{P_{i,j}d_{i,j}^{-\alpha}|h_{i,j}|^2}{G_i^{-1}G_j^{-1}q\sigma^2}}
\end{equation}

Total outage probability can be written as: 
\begin{equation}
    P_{out}^{all} = 1 - (1-P_{out}^u)(1-P_{out}^d)~\times~  \prod_{\forall i,j \in N~\&~i \neq j} (1 - P^{i,j}_{out}) ~\times~S_{(i,j)}  
\end{equation}

We define a new metric resilience score over each path, which for is a combination of outage probability and system queue conditions. 

\begin{align}
R^{all} &= \omega_1( 1 -  P_{out}^{all}) \\
&\quad + \omega_2 \max_{\substack{i,j \in N \\ i \neq j}} 
\left( \max(1 - q_i, 1 - q_j) \cdot S_{(i,j)} \right)
\end{align}

\subsection{Objective}
Our goal is to deliver data packets from source gateway $s\in G$ to receiver destination gateway $d \in G$ through satellite constellation. Our objective is to minimize the latency while minimizing the inherent uncertainties of the network. Morever we optimize compute and offloading energy while also considering communication overhead. The complete problem can be formulated as follows. 

\begingroup
    \allowdisplaybreaks
        \begin{align*}
            &\underset{ R_{g,d}}{
                \min } :
                  \sum_{p=1}^{|P_{gd}|}D_p(g, d), \forall g, d \in \mathbb{G}, g \neq d 
                 \\
            &\text{Subject to:} \\
            C1&: f_{i, j} \leq R(i, j) \forall i, j \in N
            \\
            C2&: \sum_{e_j \in \varepsilon_{i,s}} f_{j,i} = \sum_{e_k \in \varepsilon_{i,s}} f_{i,k} \forall i \in \mathbb{S} 
            \\
            C3&:  f_{i,i} = 0 \forall i \in N
            \stepcounter{equation}\tag{\theequation}\label{eq:problem_def}
            \\
            C5&: \bigg( 1 - (1-P_{out}^u)(1-P_{out}^d) \\  & \times  \prod_{\forall i,j \in N\&i \neq j} (1 - P^{i,j}_{out}) \bigg) \leq P_{out}^{threshold}   
            \\
        \label{opt-eq}
        \end{align*}
    \endgroup

\section{Proposed Solution Approach}

In this work, we utilize a \gls{drl}-based approach to address the formulated problem. Our methodology is inspired by the framework presented in \cite{1__lozano2025continual}. Specifically, we consider a Kepler, Starlink, and OneWeb constellation comprising $O$ orbital planes, with $S$ satellites deployed in each orbit. On the ground, we assume the presence of $G$ ground terminals, where each terminal serves multiple users.

Each ground terminal handles data generated by its associated users in real time, resulting in continuous traffic generation across the network. In particular, each user may transmit data to another user connected either to the same or a different ground terminal. If both the source and destination users are associated with the same ground terminal, the data is delivered locally without traversing the satellite constellation. Otherwise, the data is forwarded to the \gls{leo} satellite network and subsequently routed to the destination ground terminal.

From a system-level perspective, this interaction results in traffic exchange among all ground terminals. To model this behavior, we consider two types of traffic generation patterns: (i) uniform traffic generation, where each ground terminal produces an equal amount of data, and (ii) population-based traffic generation, where the data generation is proportional to the number of users connected to each ground terminal. These traffic patterns introduce additional background load on the \gls{leo} constellation during the transmission process between the source and destination \cite{1_2__soret2024q}.

We employ a \gls{drl} framework to train a model that can optimally route data from a source to a destination based on the current traffic conditions in the \gls{leo} satellite constellation.
To formulate the problem as a Markov Decision Process (MDP), several key components must be defined. First, we define the state representation. The state includes the coordinates of the current satellite, the coordinates of its four neighboring satellites, the queue levels associated with each neighbor (across their respective queues), and the coordinates of the packet destination. This state formulation enables the \gls{drl} agent to make routing decisions based on both network congestion and spatial information.

The action space is defined in terms of the next-hop routing decision. Specifically, each agent selects one of the neighboring satellites as the next hop, corresponding to four possible directions: upward, downward, rightward, or leftward.
The reward function is formulated as a combination of four key metrics. First, it accounts for the queueing delay at the current satellite prior to forwarding. Second, it captures the reduction in distance toward the destination achieved by the selected action. Third, a penalty is imposed to discourage revisiting previously traversed nodes, thereby avoiding routing loops. Finally, the reward incorporates the resilience associated with the selected path, promoting more reliable routing decisions.

We adopt a Double Deep Q-Network (DDQN)-based \gls{drl} algorithm to solve the considered problem. The proposed framework consists of a neural network that approximates the action-value (Q) function, a target network used to stabilize training by periodically updating its weights, and a replay memory buffer for experience storage and sampling.
Initially, we train a global Q-network to learn routing policies across the entire satellite constellation. This centralized training enables the model to capture the overall traffic dynamics and network conditions. Once the global model converges, it is deployed on individual satellites, where each utilizes the trained model to make local routing decisions. Furthermore, we incorporate online learning at each satellite to continuously update the model based on local observations, thereby improving decision-making in dynamic network conditions.

We incorporate a resilience score for each satellite link as a measure of its reliability. This consideration is particularly important in the presence of potential jamming attacks, which may render multiple satellites or inter-satellite links temporarily unavailable. The resilience score represents the confidence that a given link will remain available for the duration required to forward a packet toward its destination. This metric is integrated into both the reward function and the input state representation of the proposed model. Specifically, we extend the state space by introducing four additional features corresponding to the resilience score of the four connected links, where values range from 0 (highly vulnerable) to 1 (highly reliable). The action space remains unchanged, with the model selecting the next-hop link for packet forwarding. The reward function is extended with a fourth component, the resiliency score, which provides a reward proportional to the resilience of the selected route. The simulation environment is configured based on \cite{1__lozano2025continual} and \cite{29__abdulkarim2024multi}. The detailed configurations are summarized in Table~\ref{table_simulation_parameters}.
\begingroup
\begin{table}[]
\centering
\caption{Simulation parameters.}
\label{table_simulation_parameters}
\renewcommand{\arraystretch}{1.0} 
\resizebox{0.88\linewidth}{!}{
\begin{tabular}{p{1.4in}    p{0.5in}    p{1 in}   }
\toprule
{\textbf{Description}} & {\textbf{Parameter}} & {\textbf{Value}}  \\ \hline
Satellite constellations &  & Starlink shell 1\\
Orbital planes & $O$ & 72 \\ 
Total satellites & $S$ & 1584\\ 
Altitude & $h_0$ & 550 km \\ 
Satellite architecture &  ~~ & Walker Delta \\ 
Num. of ground terminals&  & 200\\ 
Power budget for each ground terminal link & $P_{i,j}$ & 10 W\\ 
Power budget for a satellite link& $P_{i,j}$ & 20 W\\ 
Bandwidth & B & 500 MHz\\ 

Noise & $\sigma^2$ & -174 dBm/Hz\\
Speed of light &$c$ & $3 * 10^8$ m/s  \\
Radius of Earth & $r_e$ & 6371 Km\\
Antenna gain &$G_i^{d}, G_j^{u}$ & 60 DB \\
Path loss exponent & $\alpha$ &  2 \\
Nakagami-m parameter& & 2\\
Frequency & $f_c$ & $30 * 10^9 Hz$ \\
Packet size &$B$ & 64 kb \\

\gls{drl} architecture && DDQN \\
Policy iteration &  - & Epsilon Greedy policy \\
Training iterations &  -  & 100000 \\
Replay memory &  -  & 2000 \\
\gls{drl} batch size & - & 128 \\
Epsilon start & - & 0.99 \\
Epsilon end & - & 0.1 \\
Epsilon decay rate & -  & 1000 \\
Loss function & - & Huber loss \\
Optimizer  & - & Adam \\
Optimizer learning rate & - & 0.0001 \\
Policy neural network & - & DNN with three layers \\
Target neural network & - & DNN with three layers \\

\bottomrule
\end{tabular}
}\vspace{-15pt}
\end{table}
\endgroup

\section{Performance Evaluation}

In this work, we adopt a \gls{drl} framework for routing decisions. The primary motivation for selecting \gls{drl} over traditional routing algorithms lies in its computational efficiency during deployment. Classical shortest-path algorithms, such as Dijkstra’s algorithm and Bellman–Ford algorithm, incur computational complexities of $O(E + Vlog(V))$ or $O(v^2)$ in dense graphs and $O(VE)$, respectively where $V$ and $E$ denote the number of nodes and edges. These complexities become prohibitive in large-scale satellite constellations with rapidly changing topologies. Satellite networks are inherently dynamic, with link states potentially changing on the order of milliseconds, necessitating near real-time routing decisions. In contrast, once trained, the \gls{drl} model performs inference through a forward pass of the neural network, enabling routing decisions in constant time, i.e., $O(1)$. This makes \gls{drl} particularly well-suited for such environments, as it avoids repeated global recomputation. Consequently, the proposed approach enhances the stability and reliability of the network by enabling fast and adaptive routing in the presence of frequent topology changes. Our observations indicate that the developed solution achieves performance comparable to \cite{1__lozano2025continual} in terms of data blocks lost, where both approaches report negligible packet loss under their respective definitions. 

We evaluate the performance of the proposed framework against three baseline approaches at Starlink Shell 1. We consider a \gls{sarsa}-based method trained in a centralized manner using complete constellation information and then copied on each node to get better performance, serving as a learning-based benchmark. In addition, we include the traditional routing baseline Dijkstra’s algorithm to provide a comparison with the classical method.

\begin{figure}
\centerline{
    \includegraphics[width=3.4in, clip]{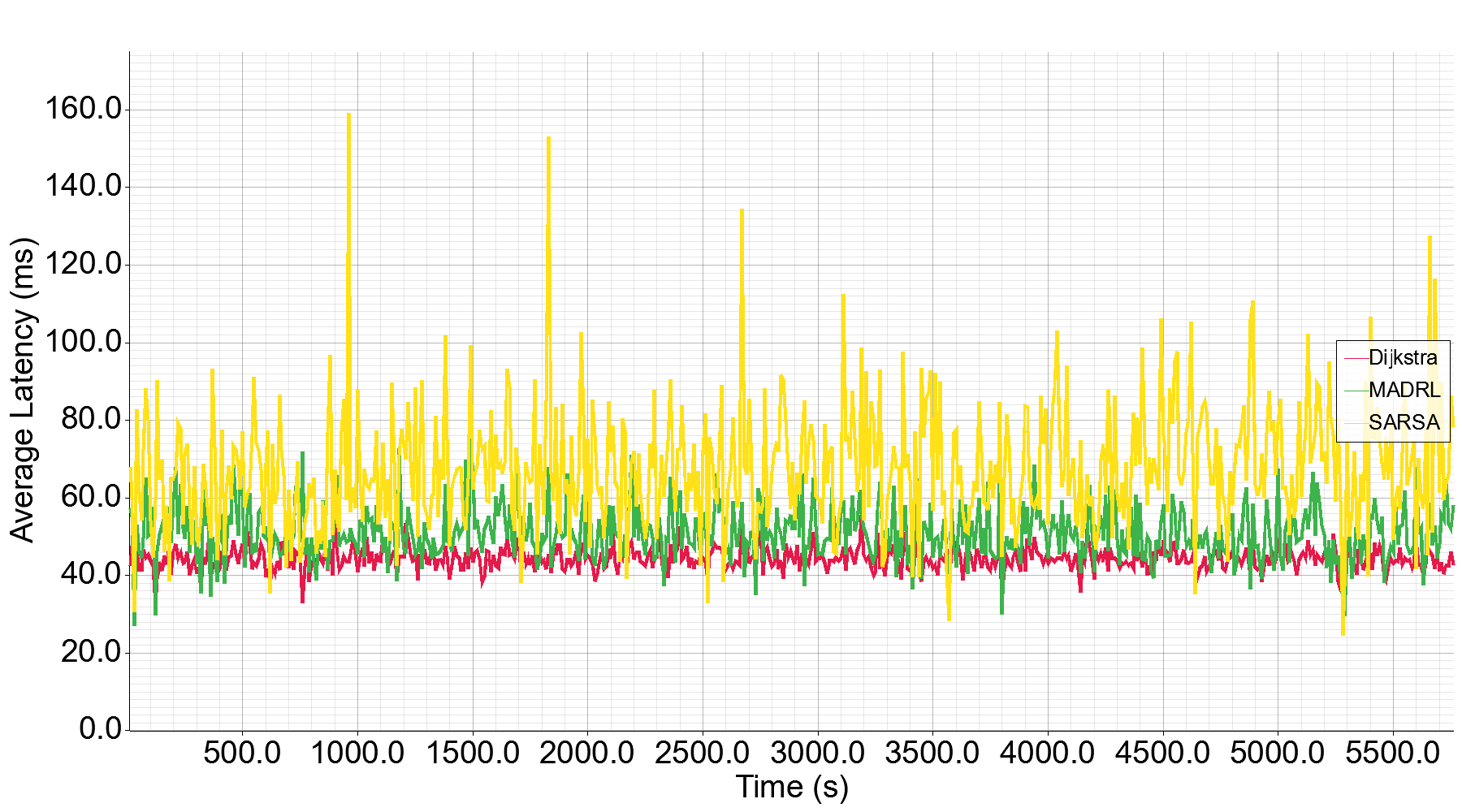}
    }
    \caption{Latency comparison of Dijkstra, \gls{sarsa}, and \gls{madrl}.}
\label{fig_latency_dijestra_vs_madrl}\vspace{-15pt}
\end{figure}

Fig.~\ref{fig_latency_dijestra_vs_madrl} presents a comparison of the packet delivery latency obtained using the Dijkstra’s algorithm and the proposed \gls{madrl} approach. In the baseline implementation, Dijkstra’s algorithm utilizes link latency as the edge weight to construct routing tables, enabling each satellite to forward packets along the minimum-latency path. However, this approach does not incorporate queue state information in its decision-making process. To ensure a fair comparison and isolate latency performance, the queue capacity in the simulation is set to 1~Gb/s, effectively preventing queue overflow and allowing Dijkstra’s algorithm to operate under near-ideal conditions. Under the same simulation setup, the proposed \gls{madrl}-based routing is evaluated, and the corresponding latency is measured. The results indicate that the average latency achieved by \gls{madrl} is 49.31 ms, compared to 38.54 ms for Dijkstra’s algorithm. \Gls{sarsa} algorithm has the worst performance as compared to the other two approaches, with an average latency of 59.28 ms. While Dijkstra’s algorithm shows lower latency under these controlled conditions, it does not explicitly account for queue dynamics. In contrast, the proposed \gls{madrl} framework and \gls{sarsa} incorporate queue backlog information into their decision process, enabling it to mitigate packet drops in more realistic network scenarios.

\begin{figure}
\centerline{
    \includegraphics[width=3.2in, clip]{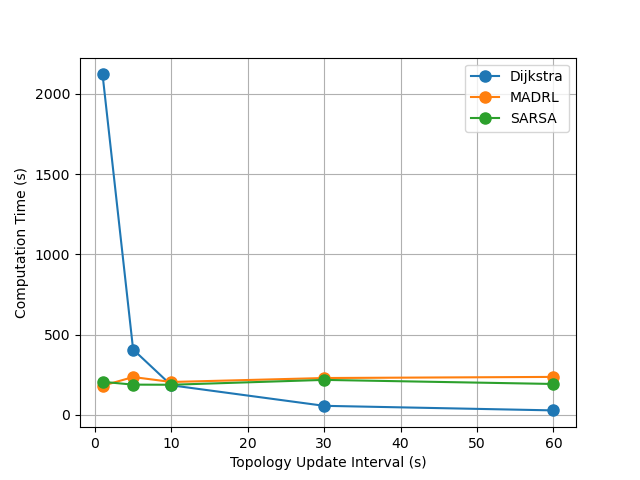}
    }
    \caption{Aggregate decision-making time for each routing algorithm.}\vspace{-12pt}
\label{fig_decisioncost_dijestra_vs_madrl}
\end{figure}

In Fig. \ref{fig_decisioncost_dijestra_vs_madrl}, we show the impact of the cost associated with routing decision computation across the entire satellite constellation aggregated. Due to the dynamic nature of the constellation, Dijkstra’s algorithm must repeatedly recompute routing paths to adapt to topology changes. Even when deployed on a high-performance ground base station, this frequent recomputation, combined with the overhead of disseminating updated routing tables to all relevant satellites, results in a significantly higher decision cost. \Gls{sarsa} and \gls{madrl} are deployed on each satellite, therefore the cost of getting a decision for each one is forward inference of their neural networks. As we decrease the frequency of updates based on changing topology, we see a decrease in the computation time for Dijkstra. This is because most of this overhead time is utilized to compute the global state in Dijkstra's algorithm, which is then utilized to update local routing tables at each satellite. We observed that after increasing the update frequency for Dijkstra to more than 10 s, the cost of decision making becomes less then of \gls{madrl} and \gls{sarsa}, but this causes other unintended side effects. 

\begin{figure}
\centerline{
    \includegraphics[width=3.2in, clip]{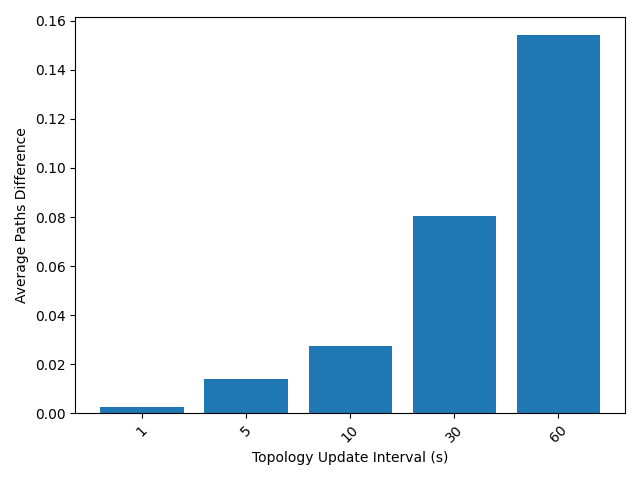}
    }
    \caption{Aggregate path changes for the Dijkstra algorithm.}
\label{fig_dijestra_lost_performance}\vspace{-18pt}
\end{figure}

Fig. \ref{fig_dijestra_lost_performance} shows the average path changes in percentage as the route recalculation frequency of Dijkstra-based routing increases. While less frequent recalculations reduce decision cost, they come at the cost of optimality, as rapid topology changes make previously optimal paths suboptimal. Some of the paths that were available before are no longer available. This leads to increased overall latency despite faster decision updates. In contrast, both \gls{madrl} and \gls{sarsa} rely on pre-trained models and perform only inference. As a result, their performance remains largely unaffected by topology variations, enabling more stable and consistent latency without the need for continuous route recomputation.

\begin{figure}
\centerline{
    \includegraphics[width=3.2in, clip]{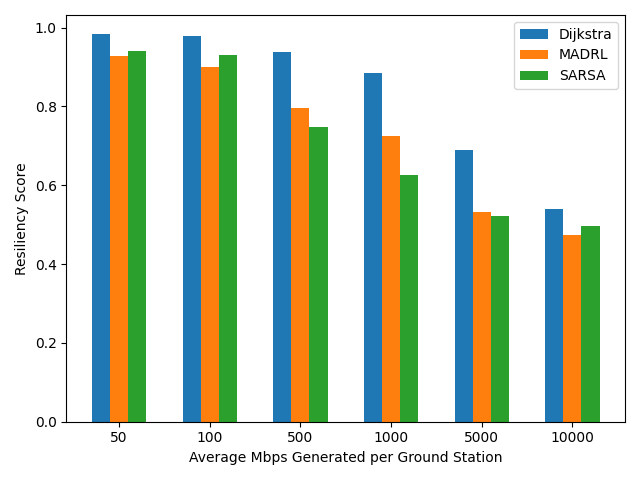}
    }
    \caption{Comparison of resilience scores across different routing algorithms under varying traffic conditions.}\vspace{-15pt}
\label{fig_resilience_score}
\end{figure}

Fig.~\ref{fig_resilience_score} shows the average resilience score across all network links over the entire simulation, under varying levels of background traffic, expressed as the average data generated by each base station. As observed, Dijkstra achieves the highest resilience score. This is primarily due to its global view of the network, allowing it to make routing decisions based on complete constellation information and overall traffic conditions. In contrast, the proposed approach exhibits comparatively lower resilience. Although each agent has access to its local queue state as well as the queue conditions of its neighboring nodes and attempts to make optimal routing decisions, it lacks a global view of the network. This limited observability results in slightly reduced resilience performance for both \gls{madrl} and \gls{sarsa}. Furthermore, the resilience scores of \gls{sarsa} and \gls{madrl} are comparable, as both approaches follow a distributed decision-making architecture.

\section{Conclusion}
In this work, we investigated routing in \gls{leo} satellite constellations under dynamic topology and traffic conditions. We incorporated end-to-end latency, queue dynamics, and a resilience score into the system model to provide a more realistic representation of the network environment. Unlike traditional approaches (e.g., Dijkstra-based), the proposed formulation accounts for queue backlogs as well as real-time changes in network topology. We developed a queue-aware, \gls{madrl} framework, in which each satellite operates as an independent agent making routing decisions based on local observations. This distributed design enables timely decision-making while allowing the agents to continuously learn network traffic patterns, thereby mitigating congestion and improving resilience under high system utilization. The experimental results validate the effectiveness of the proposed MA-DRL approach, which sustains stable performance with lower overhead by proactively managing queue states and link resilience. Future work will investigate advanced learning techniques and hybrid learning–traditional formulations for improved efficiency and robustness.


\ifCLASSOPTIONcaptionsoff
  \newpage
\fi

\bibliographystyle{IEEEtran}
\bibliography{references}



\end{document}